# Shape anisotropy enhances the elasticity of colloidal gels through mechanisms that act multiplicatively

Peng-Kai Kao[†], Michael J. Solomon[†,*], and Mahesh Ganesan[†,*]

[†]Department of Chemical Engineering, University of Michigan, Ann Arbor, Michigan 48109, USA

\*Corresponding Authors:

Prof. Michael J. Solomon

Address: North Campus Research Complex, Building 10 – A151, 2800 Plymouth Road, Ann Arbor, MI 48109

Phone: 734-764-3119

Email: mjsolo@umich.edu

Dr. Mahesh Ganesan

Address: North Campus Research Complex, Building 20 –106W, 2800 Plymouth Road, Ann Arbor, MI 48109

Phone: 734-546-0210

Email: maheshg@umich.edu

**Abstract**

The physical gelation of colloids produces elastic structures that are commonly used to stabilize complex fluids in multiple industries. However, the ability to control the level of elastic modulus of these materials is limited by the universality of the arrested spinodal decomposition mechanism that governs their formation. We demonstrate that manipulating particle shape can yield very large shifts in gel elasticity relative to this universal behavior, and that these changes are generated by a set of independent mechanisms that act multiplicatively. Modeling demonstrates that this efficient generation of elasticity is the consequence of the combined effects of shape anisotropy on the hierarchical fractal microstructure of the gel network, the anisotropy of the attractive interparticle pair potentials, and the volumetric compactness of the fractal cluster. The findings show how and why particle anisotropy can be exploited to generate elasticity at ultra-low colloidal volume fractions in this important class of soft matter.

**Introduction**

Colloidal gels are tenuous, sample spanning structures that exhibit soft, solid-like rheology such as finite elastic modulus and yield stress[1]. They are commonly used to impart quiescent elasticity to complex fluids such as paints, agrochemicals, and cosmetics[2,3]. Gels are formulated by quenching particles through strong, short-range attractive pair potentials[4]. However, especially at low particle volume fractions ($\phi \lesssim 0.1$) – where industry often seeks to operate due to cost – rheological control of gels is limited by the universality inherent to colloidal gelation[5,6]. Specifically, studies have established that gelation occurs through a process of arrested spinodal decomposition[4]. At these dilute concentrations, this mechanism of spinodal decomposition proceeds by diffusion limited cluster aggregation (DLCA). DLCA at finite volume fraction generates a microstructure of uniformly distributed, conjoined fractal clusters[1,4].

Within the clusters the fractal dimension ($d_f$) of the DLCA gel microstructure is a universal value, $d_f \approx 1.85$.[5] The only other microstructural parameter, the cluster size, is specified by the gel volume fraction. Gel rheology is consequently tunable only through particle volume fraction, $\phi$ and interparticle bond strength, the latter of which is quantified as the bond spring constant, $\kappa_0$ [7]. By drawing from data from multiple researchers, Romer et al show that in the range $10^{-3} \leq \phi \leq 10^{-1}$, the $\phi$ dependence of the linear elastic modulus, $G'$ of colloidal gels consisting of spherical monomers collapses onto a universal curve when normalized by particle size[8].

The universality of the mechanism of arrested spinodal decomposition thus constrains opportunities to design the elasticity of colloidal gels. Even volume fraction and interparticle bond strength are often unavailable to formulators as design parameters because they are otherwise deployed to achieve the gel's intended function. For example, factors that mediate attractions such as salinity, solvency, and additive concentration are often used to support the activity of a formulation. Technologically, it would be useful to engineer efficient gels with high $G'$ at low volume fraction since this combination is effective from the point of view of both sustainability and cost. Here we show how and why shape anisotropic particles step outside the constraints imposed by the universality of arrested spinodal decomposition. In doing so, we reveal and characterize the ultra-efficient generation of elasticity in colloidal discoid gels – at volume fractions up to 15 times less than spherical gels. The reason for this shift – which is equivalent of up to a factor of $10^2$ increase in elastic modulus at fixed volume fraction – is that anisotropy affects three features of the gel network – fractal geometry, interparticle bond strength and cluster compactness – in ways that combine multiplicatively to generate large effects on gel mechanics.

There are prior indications that shape anisotropic colloids produce fractal objects with unusual microstructural and microrheological behavior. For instance, DLCA of colloidal rods yielded networks with $d_f$ that had a strong dependence on particle aspect ratio, deviating from the universal value of $\approx 1.85$ found for colloidal spheres[9]. The localized microdynamics of rod aggregates were reported to show an unforeseen transition from floppy to brittle behavior that departed from microrheological theories of cluster dynamics[10]. Other early studies showed that aggregate networks of anisometric colloids exhibit higher interfacial modulus[11] and yield stresses[12,13] than the case in which the primary constituents were simply spheres. However, the specific mechanism by which shape anisotropy mediated these different elastic states is not known. Speculations as to the mechanism have included non-central forces[10,13], anisotropy in excluded volume[9] as well as an increase in the number of contacts per particle[10,14] due to the asphericity of particle shape. Moreover, although anisotropy effects on gelation boundaries have been discussed[15], the specific shifts in gel elastic modulus due to anisotropy have not been quantified.

We expand upon the above findings in the context of gel design to explore the idea that particle shape anisotropy can generate microstructures that deviate from the universal features of sphere gels, enabling greater control over gel rheology. We use colloidal discoids as a model system because recent progress in fabrication techniques yield uniformly sized colloids in quantities sufficient for rheometry[16]. The discoid shape is furthermore common in engineered[17] and living systems[18]. Moreover, studies investigating gelation of laponite platelets – nanometer sized clay minerals – show that the discoidal geometry of the monomer gives rise to intriguing microstructures – such as a 'house-of-cards' assembly – leading to an increase in their elastic rheology[17].

In the present work, we observe that the $\phi$ dependence of $G'$ in discoid gels exhibits a remarkable shift from the universality of sphere gels as aspect ratio, $r$, is changed. In addition to a shift of the modulus curve to very low volume fractions, we also find that the power-law exponent describing the $\phi$ dependence of $G'$ changes with $r$. We investigate the microstructural origin of these phenomena by characterizing network connectivity through confocal microscopy and morphological image analysis. These independent measures of microstructure are inputs into the well-known theory for the elasticity of fractal cluster gels[19]. We find that the fractal dimension and shape of the elastic backbone in discoidal gels differs from the universal values for spheres; these differences explain the $r$ dependence of the $G' - \phi$ power-law exponent. The anisotropy in pair potential interactions further contributes an aspect ratio dependent shift in the volume fraction dependence of the elastic modulus. The remaining shape dependence is captured by a single $r$-dependent prefactor, the dependence of which suggests an additional effect of cluster compactness on rheology. Remarkably, each of these effects aligns with the others, generating the large multiplicate effect of shape anisotropy on gel rheology.

This work reveals a simple, efficient means for expanded design and control of the rheology of colloidal gels. In addition to its direct utility in producing and using discoid gels, the analysis reveals why shape has the effect on gel rheology that it does and suggest avenues through which other kinds of anisotropic shapes and interactions might be used to affect the underlying variables of hierarchical fractal structure, spring constant, and cluster compactness.

Our study therefore motivates further research on the capabilities afforded by anisotropic building blocks to self-assemble into colloidal gels with previously unavailable structure-property relationships.

**Results**

**Colloidal gels assembled from discoids of different aspect ratios**

Discoids used in this study are prepared by thermo-mechanical compression of spheres[16]. The seed microspheres (diameter $2A = 1.00 \pm 0.02$ µm) were embedded in a polymer film and heated above their glass transition temperature in a bench-top press (c.f. Methods, Supplementary Figure 1). Uniaxial compression at pressures of 750, 2800, and 4200 kPa was applied, resulting in discoids with aspect ratio, $r = (B/A) = 0.42 \pm 0.01$, $0.20 \pm 0.01$ and $0.09 \pm 0.03$ respectively. The major axes for these particles are $2A = 1.34 \pm 0.02$ µm (for $r = 0.42$), $1.72 \pm 0.01$ µm (for $r = 0.20$) and $2.23 \pm 0.04$ µm (for $r = 0.09$); the corresponding minor axis are $2B = 0.56 \pm 0.01$ µm, $0.35 \pm 0.01$ µm and $0.21 \pm 0.01$ µm. Figure 1(a)-(d) show scanning electron micrographs of the seed spheres and the as fabricated discoids. Insets in Figure 1(b)-(d) provide an orthogonal view; the shape uniformity and successive flattening of the spheroidal shape of the particle with decreasing aspect ratio are apparent.

Particles were suspended in a density-matched mixture of $H_2O$-$D_2O$ and gelation was induced by addition of 10mM $MgCl_2$[20]. The salt screens surface charges promoting short-range van der Waals attractions that drive gelation through a DLCA process. Figure 1(e)-(l) compares maximum projections of the three-dimensional structure of the self-assembled colloidal gels. Researchers working with gels formed by discotic clay minerals have described the microstructure as being a 'house-of-cards' network; however, we here see little evidence for such structures in Figure 1[17]. Instead, a hierarchical, self-similar, fractal microstructure is apparent for both the sphere ($r = 1.00$) and discoid ($r = 0.09 – 0.42$) gels. The heterogeneous, inter-connected microstructures that span the image space at intermediate length scales (Figure 1(e)-(h)) are seen to persist even at smaller length scales (Figure 1(i)-(l)). This description is consistent with the well-known fractal cluster architecture of dilute colloidal gels[7]. In addition, the gels formed by discoids possess a microstructure (Figure 1(j)-(l)) that is interspersed with both condensates – aggregates with no orientational ordering – and strands – particles with a face-to-face orientational alignment. This mixture of partially ordered and disordered aggregates is similar to the configuration reported by Hsiao et al. for discoids assembled through depletion forces[21].

**Elasticity shift of discoid gels**

In the volume fraction range studied here, both sphere and discoid gels exhibit similar functional dependence of $G'$ on oscillatory strain ($\gamma$): a linear regime with plateau elasticity at low strains ($\gamma < 0.003$) followed by an onset of non-linearity and a rapid drop in modulus with increasing strain amplitude (Figure 2(a)-(b), Supplementary Figure 2). The $\tan(\delta)$ – defined as the ratio of viscous to elastic modulus – is about 0.20 in the linear regime and increases to values $> 1$ in the non-linear regime (inset in Figure 2 and Supplementary Figure 2). The finite $G'$ and $\tan(\delta) < 1$ together indicate a solid-like, elastically dominated response, characteristic of physical gels. At higher strains, the rapid loss of elasticity represents a fluidization transition[22].

Extreme differences between discoidal and spherical gels are seen by plotting the dependence of their linear $G'$ on volume fraction (Figure 3). First, we observe elastic rheology in discoid gels at a volume fraction that is more than an order of magnitude lower than spheres. For instance, sphere gels display $G' = 0.2$ Pa at $\phi = 1.5 \times 10^{-2}$ while gels consisting of $r = 0.20$ and $r = 0.09$ discoids already exhibit similar elasticity at $\phi = 2.3 \times 10^{-3}$ and $1.1 \times 10^{-3}$ respectively. Relatedly, discoidal gels exhibit between $10^1$–$10^2$-fold higher elastic modulus than spherical gels at fixed volume fraction, with level of stiffening increasing with anisotropy. For example, at $\phi = 1.5 \times 10^{-2}$, $G' = 0.17$ Pa and 18 Pa for sphere and $r = 0.20$ discoid gels, respectively. At $\phi = 7.5 \times 10^{-3}$, gel modulus increases by 20-fold as discoid anisometry increases from $r = 0.42$ to 0.09.

Second, we find that the exponent, $\Pi$, describing the $\phi$ dependence of $G'$, $G' \sim \phi^\Pi$, decreases as $\Pi = 3.53 \pm 0.16$, $2.93 \pm 0.10$, $2.52 \pm 0.05$ and $2.15 \pm 0.15$ for $r = 1$, 0.42, 0.2 and 0.09 respectively. These differences in the power law behavior are statistically significant (one-way ANOVA yields $p = 0.0048$). The value $\Pi = 3.53$ for the spheres agrees with previous reports, which average $\Pi = 3.51 \pm 0.15$ [7,8]. The slower growth in elastic modulus with volume fraction for the anisotropic gels – as represented by the decreasing power law – is in itself noteworthy and agrees with the shear rheology of rod[12,13] and platelet clay gels[23]. In addition to this feature, the power law curve of the elastic modulus itself shifts to progressively lower volume fractions with aspect ratio, culminating with a shift of $22.4 \pm 5.5$ at the lowest discoid aspect ratio. Figure 3 therefore shows that the shifts in pre-factor and exponent combine to generate very large absolute effects of shape anisotropy on gel rheology.

**Model to investigate mechanism of the elasticity shift**

To develop an explanation for the aspect ratio dependence of the elasticity shift, we apply a well-known theory for dilute colloidal gel rheology developed by Shih et al[19] and further extended by de Rooij et al[24] and Krall and Weitz[7]. Briefly, in this microelastic model, the gel is described as a uniform packing of fractal clusters. The characteristic cluster size, $R_c$, is set by the particle volume fraction, $R_c \sim V_p^{1/3} \phi^{-1/(3-d_f)}$, where $V_p$ is the hard particle volume[7]. The gel elastic modulus is then determined by the spring constant, $\kappa(R_c)$ of these clusters: $G' \sim \kappa(R_c)/R_c$. The spring constant of the cluster is determined by the bending rigidity of its skeletal backbone – a fractal chain of particles spanning the length of the cluster[25]. Relatedly, $\kappa(R_c)$ is given as $\kappa(R_c) \sim \kappa_0 R_c^{-\beta}$, where $\kappa_0$ is the bending constant of the bond between a pair of particles within the cluster and $\beta$ is determined by the fractal structure of the cluster spanning backbone. Specifically, $\beta = 2\epsilon + d_B$, where $d_B$ is the backbone fractal dimension and $\epsilon$ represents the anisotropy of the backbone[24]. The model prediction for $G'$ is then given as:

$$G'(\phi, r) = f \kappa_0 V_p^{-\frac{1}{3}} \phi^{\frac{1+\beta}{3-d_f}} \quad (1)$$

Equation (1), applicable in the dilute limit ($\phi < 0.1$)[24], describes a power-law dependence of $G'$ on $\phi$. The pre-factor of the scaling is determined by three parameters. They are: $\kappa_0$, $V_p$ and, $f$ – a proportionality constant. The power-law exponent depends on $d_f$ and $\beta$, which together describe the hierarchical microstructure of the gel. The characteristic length scale in equation (1) has been taken as $V_p^{1/3}$, rather than the particle radius, in anticipation of our applying the model to anisometric particles[14,26].

The model is generally applicable to gel materials with fractal cluster structure; it does not prescribe the shape of the gel's constituent particles [8]. As a consequence, it has been widely used to model the rheology of gels formed from spheres (e.g. latex colloids[7,8,27], mineral

particles[19,28]), spheroids (e.g. carbon nanotubes[29], boehmite ellipsoids[30], starch granules[31]), polymeric chains (e.g. proteins[32], polysaccharides[33], fat molecules[34]) and patchy particles[35].

Calculation of the single-bond rigidity, $\kappa_0$, has been addressed by many authors. For long backbones, Kantor and Webman show that deformation occurs through angular rotation of backbone bonds[25]. Such resistance to bond bending implies the existence of non-central forces. These forces, especially for strong pair potentials, have been described as arising due to surface roughness, surface deformation, or steric stabilization[24,36]. Dinsmore et al. showed that fluctuations in fractal clusters were consistent with bond bending dynamics for strongly bonded gels[36]. Pantina and Furst successfully modeled $\kappa_0$ by applying Johnson-Kendall-Roberts (JKR) theory; the contact area generated by adhesive interactions supported bond flexure[37]. In this scenario, the bending rigidity tracks the strength of pair potential interactions, as described, in the present system, by the Derjaguin-Landau-Verwey-Overbeek (DLVO) potential. Pantina's calculations further suggest that computing $\kappa_0$ from thermal fluctuations of the colloid pair separation represents an upper bound to the actual bending constant[38]. Similar approaches have been used to model the steady shear viscosity of aggregating latex[39], the yield rheology of polystyrene and alumina gels[40], and the elasticity of dense depletion gels[41]. Here we pursue this approach to compute $\kappa_0$ from the thermal fluctuations of the colloid pair bond separation, weighted by the probability of different orientations for discoids.

Numerical values for the parameters in equation (1) are well known for DLCA sphere gels: $d_f \approx 1.85$, $\beta \approx 2.80$ and $f \approx 1 - 3$ [7,8,27,42]. However, these parameters may vary when the interparticle interactions and shape of constituent monomers differ from those of simple spheres.

For example, from measurements of the dynamic structure factor, Mohraz and co-workers found that gels consisting of rod-like particles displayed variations in both $d_f$ and $\beta$ with aspect ratio[9,10]. Simulations by West et al[43] showed that introducing angular rigidity in the gel backbone resulted in values of $\beta < 2$. Through theoretical fits to experimental data, Laxton and Berg reported a pre-factor of the modulus scaling with volume fraction that increased up to $\approx 50$ for discotic clay gels[44]. Therefore, each of the parameters in equation (1) – $d_f, \beta, f$ and $\kappa_0$ – may be aspect ratio dependent, a possibility we now explore. In the subsequent sections, we report independent, confocal microscopy derived, measurements of $d_f(r)$ and $\beta(r)$, theoretical calculations of $\kappa_0(r)$ and a least-squares determination of $f(r)$. Finally, using equation (1), we discuss how aspect ratio induced changes to these parameters act as multiplicative contributions to produce large shifts in discoid gel rheology, as reported in Figure 3.

**Fractal dimension of the gel network**

We measure $d_f$ from confocal image volumes using box-counting analysis, which utilizes spatial intensity data to quantify the dimensionality of the structure[45]. This method has previously been applied to quantify the microstructure of colloidal sphere gels[46,47], albumin gels[48], plasma protein gels[49] and soot aggregates[50]. The method is well-suited to the discoidal shape studied here. Briefly, the number of cubes $N(L)$ of size $L$ required to cover the gel are computed (Figure 4(a)) at different cube sizes, yielding $d_f$ from the power-law $N(L) = c_1(L/I)^{-d_f}$, where $I$ is the image size and $c_1$ is a proportionality constant (c.f. Methods). Figure 4(b) shows the resulting log-log plot. The fitted lines corresponding to all $r = 0.09 - 1.00$ gels have power law slope of magnitude smaller than three, confirming the fractality of these structures.

Figure 4(c) shows that increasing particle shape anisotropy causes an increase in gel $d_f$. For sphere gels ($r = 1$), $d_f = 1.86 \pm 0.02$, is consistent with the DLCA value of $d_f \approx 1.85$. For discoid gels ($r < 1$), the value increases up to $d_f = 2.04 \pm 0.02$ for the lowest aspect ratio ($r = 0.09$). This result is consistent with prior measurements for gels of rodlike particles ($r > 1$), which also reported an increase in gel $d_f$ with increasing monomer anisotropy[9]. In Figure 4(b), the curves display an upward shift with aspect ratio, indicating an increase in $c_1$ (inset Figure 4(c)).

The increase in network dimensionality indicates a deviation from the universality of DLCA aggregation[5]. Higher $d_f$ indicate that discoids form denser cluster than spheres. The reason for the increased fractal dimension is attributed to the anisotropy in excluded volume as per simulations reported by Mohraz et al[9]. Physically, the increase in the proportionality constant $c_1$ as discoid anisotropy increases (aspect ratio decreases) indicates that the discoid fractal structures are more space filling relative to spheres at any length scale and controlling for fractal dimension. For instance, for a given box-size, the number of boxes required to completely fill the gel microstructure is seven-fold higher for the case of $r = 0.09$ discoid gels than that required for sphere gels. The dependence of $c_1$ on $r$ thus shows that DLCA fractal structures produced from discoids are more compact than spheres.

**Backbone dimension and isotropicity**

We extract the gel backbone from confocal micrographs using skeletonization – a sequential thinning process that produces a voxel-thick skeletal backbone that encodes the topography of the gel (c.f. Methods). It has been applied to measure backbone tortuosity of pNIPAm colloidal gels[51], nanoparticle aggregates[52] and osteocyte networks[53]. The backbone dimension, $d_B$ is obtained from the relation between the length of the shortest path, $l$, between branch points on the skeleton and the end-to-end distance, $r_E$ between them: $l = c_2 r_E^{d_B}$ [42,54]. The backbone anisotropy, $\epsilon$ is given as $r_\perp^2 = c_3 r_g^{2\epsilon}$ where, $r_\perp$ is the radius of gyration of the shortest path projected onto a plane perpendicular to its end-to-end axis and $r_g$ is its center-of-mass radius of gyration[55,56]. de Rooij et al identify the following limiting cases for the backbone shape parameters: $\epsilon = 1$ and $d_B = 5/3$ applies to an ideal self-avoiding isotropic chain, while $\epsilon = 0$ and $d_B = 1$ applies to a straight chain[24]. Here, $c_2$ and $c_3$ are proportionality constants.

The skeletonization reveals the gel backbone microstructure as a sparse, interconnected network interspersed with voids (Figure 5(a)). Figure 5(b)-(c) shows the corresponding plots of $l$ vs $r_E$ and $r_\perp^2$ vs $r_g$ and the power-law fits to obtain $d_B$ and $\epsilon$, respectively. For sphere gels, the value $d_B = 1.26 \pm 0.03$ and $\epsilon = 0.80 \pm 0.02$ agrees well with the results of computer simulations ($d_B = 1.30$ and $\epsilon = 0.77$)[54,57] and direct measurements ($d_B = 1.20$ and $\epsilon = 0.7$)[42]. With increasing anisotropy, we find that $d_B$, $\epsilon$ (inset Figure 5(d)) and consequently the elasticity exponent, $\beta = 2\epsilon + d_B$ (Figure 5(d)) decrease, culminating at a value of $\beta = 1.66 \pm 0.08$ for the lowest discoidal aspect ratio. A similar result was observed for rod gels where a value of $\beta \approx 1.20$ for $r = 3.9 - 30.1$ was measured by dynamic light scattering[10]. The constants $c_2$ and $c_3$ are approximately unity and relatively insensitive to $r$.

The elasticity exponent $\beta = 2\epsilon + d_B$ determines the cluster spring constant[19]. Lower values of $\beta$ for increased shape anisotropy, as reported in Figure 5, are due to decreases in both $\epsilon$ and $d_B$. The results show that the backbone of the discoidal clusters becomes progressively more anisotropic relative to spheres as the discoid aspect ratio decreases. To explain the decreased $\beta$ of gels of anisometric colloids, Mohraz and Solomon made the connection between $\beta$ and backbone angular rigidity[10]. They hypothesize that, in clusters of anisotropic particles, the presence of noncentral forces arising out of additional contacts per particle limit bond rotations. This limitation leads to gel backbones with large angular rigidity. By applying the Krall and Weitz theory for the dynamics of fractal cluster gels, they reported that this rigidity results in values of $\beta$ that are lower than those for spheres. This hypothesis is supported by recent simulations showing that noncentral forces and additional contact points per particle lead to gels with less tortuous backbones[51] and higher shear modulus[58].

**Interparticle bond spring constant**

The interparticle bond strength, $\kappa_0$ is calculated from pair-potentials using the equipartition theorem, $\kappa_0 = k_B T/(\langle s^2 \rangle - \langle s \rangle^2)$ [35,36,41]. Here, $s$ is the surface-to-surface separation between particle pairs and $\langle \cdot \rangle$ represents a Boltzmann weighted average[35]. The denominator represents an ensemble average over thermal fluctuations of the interparticle separation. Discoid pair potentials are calculated using the expressions of Schiller et al[59].

For spheres, the interaction potential is isotropic; for discoids however, shape introduces an anisotropy in the attractive pair potential[21]. To highlight the interaction anisotropy accorded by particle shape, four limiting configurations of discoid pairs, namely face-to-face (F-F), edge-to-edge (E-E), edge-to-face (E-F) and edge-on-edge (ExE) and their respective pair-potentials (calculated for $r = 0.20$) are shown in Figure 6(a)-(b). Supplementary Figure 3 plots the variation of bond energies for all possible pairwise alignments. The interaction strength varies with relative orientation. F-F alignment results in the strongest potential, significantly stronger than that between two spheres – while E-E and E-F result in relatively weaker pair potentials. The stronger potential energy for F-F alignment was also observed in discoids interacting through depletion forces[21].

To calculate the interparticle bond strength between two discoidal colloids, we average over all relative orientations, weighted by their energy. That is, following Torres-Diaz et al, different orientational configurations are weighted by the Boltzmann distribution[60]. The assumption of the Boltzmann distribution is consistent with Hsiao et al[21]. In that study, the bond-angle distributions in clusters of attractive $r = 0.5$ discoids (which is intermediate to the range of $r$ studied here) displayed a higher likelihood for the energetically stronger F-F configuration[21]. Figure 6(c) shows the degree to which the Boltzmann weighted, orientationally averaged bond strength, $\langle \kappa_0 \rangle$ increases with shape anisotropy. For the lowest aspect ratio discoid, the bond strength increases by a factor of six relative to spheres. The increase in $\langle \kappa_0 \rangle$ with monomer anisotropy indicates that, on average, discoidal bonds are stronger than spheres.

**Discussion**

The parameters of equation (1), except the proportionality constant $f(r)$, are available through Figures 4–6. We first test the performance of the model and the internal consistency of the different measurements by comparing the predicted scaling exponents with those determined by mechanical rheometry. Because of the independent confocal microscopy measurements, this comparison involves no adjustable parameters. The model predicts power law exponents,

$(1 + \beta)/(3 - d_f) = 3.51 \pm 0.13$, $3.01 \pm 0.23$, $2.90 \pm 0.28$ and $2.70 \pm 0.25$ for $r = 1$, $0.42$, $0.20$ and $0.09$ respectively. These exponents can be compared to those derived from Figure 3. Both two-sample $t$-test and ANOVA comparing the experiment and model yields $p = 0.9621$, $0.7711$, $0.3068$ and $0.2059$ for $r = 1$, $0.42$, $0.2$ and $0.09$ respectively and $p = 0.1179$ across all conditions, indicating that the predicted exponents are statistically similar to the measured values. The comparison supports the application of the model to the discoidal gels; it further highlights the salience of the gel fractal and backbone dimensions to the rheological response. Figure 7(a) graphically compares the power law predictions of $G'$ using equation (1) with the rheological measurements for each of the different aspect ratios gels, with excellent agreement. The proportionality constant, $f(r)$, determined through Figure 7(a) by least squares regression, yields $f = 1.18 \pm 0.14$, $1.62 \pm 0.60$, $3.25 \pm 1.10$ and $8.55 \pm 1.22$ for $r = 1$, $0.4$, $0.2$ and $0.09$ respectively (Figure 7(b)).

With the model and characterizations in hand, we now identify the set of multiplicative factors that generate the very large shift in discoid gel rheology. At fixed $G'$, this shift is as much as a factor of 15 at the lower volume fractions of the study. Returning to equation (1) the aspect ratio dependent rheological shift is a consequence of three multiplicative factors: $\phi^{\frac{1+\beta(r)}{3-d_f(r)}}$, $\langle \kappa_0 \rangle(r)$, and $f(r)$. The first term represents particle shape induced changes to the hierarchical microstructure of the gel, as characterized by the fractal dimension and backbone topography. These quantities mediate the contribution of volume fraction to gel modulus. To assess the relative contribution of this multiplicative factor to the shift in rheology, we select $G' = 0.2$ Pa as a reference value. The choice is recommended by the fact that measurements are available at this condition for each aspect ratio, and this modulus level is not an uncommon one for practical applications of colloidal gels. At this reference value, the multiplicative factor due to hierarchical microstructure accounts for 34% of the rheological shift between $r = 1$ and $r = 0.09$. The second multiplicative contribution is the effect of anisotropic pair potential on $\langle \kappa_0 \rangle(r)$. This multiplicative factor accounts for 23% of the shift in elastic modulus between $r = 1$ and $r = 0.09$. The third term is the aspect ratio dependent proportionality constant, $f(r)$, which accounts for the remaining 43% of the shift.

The physical origin of the aspect ratio dependence of $f(r)$ is of great interest since the greatest amount of the rheological shift is embodied in this quantity. We have examined the physical factors affecting $f(r)$. At its root, these contributions involve the pre-factors that specify the quantitative relationships governing the density of the fractal network ($d_f$) and the backbone network ($\beta$, itself a function of $\epsilon$ and $d_B$). These prefactors are the constants $c_1, c_2$ and $c_3$, reported in Figures 4 and 5. Of these constants, by far the largest change is in $c_1$, and this constant was earlier identified as a measure of cluster compactness because it determines the absolute level of occupation of volumetric regions of the gel. We thus hypothesize that the third multiplicative factor affecting the rheological shift is cluster compactness. It is significant that each factor – hierarchical fractal microstructure, anisotropic pair potential, and cluster compactness – acts to shift the elastic rheology in the same direction. This alignment is the underlying origin of the enhanced effect of anisotropy on the rheological response.

The results raise a number of questions and opportunities for future research. First, although the variation in $f$ is here explained by changes to cluster compactness based as per the direct measurement of network connectivity (Figure 4,5), $f$ could also include micromechanical contributions that would be of interest to explore. For instance, Kantor and Webman showed that the bending rigidity of the cluster backbone is inversely proportional to the number of flexible bonds (pivot points[42]) in the chain[25]. If a portion of the backbone is locally rigid (i.e. locally arrested aggregates[41]), the effective spring constant would increase proportionally. This increase would in turn affect the magnitude of $f$. Therefore, probing the aspect ratio dependence of the density of angularly arrested bonds and its role in determining the macroscopic gel elasticity is suggested. An alternative micromechanical effect on $f$ could arise from the role of bond-bending mechanics in determining the single-bond rigidity $\kappa_0$ itself. As discussed earlier, Pantina and Furst showed that the bending dynamics of a linear chain of spherical colloids can be quantitatively predicted by calculating $\kappa_0$ using the JKR theory for interparticle adhesion. This approach also invokes DLVO interactions; however, in this case their contribution is mediated by the bond contact area[37]. Application of their theory for a pair of attractive polystyrene spheres suggests that the bond-bending spring constant in this case is higher than the value plotted in Figure 6 by a factor of $\approx 4$. This shift is comparable to the variability in $f$ plotted in Figure 7(b). Therefore, future work to compare the JKR prediction for the $r$ dependence[61] of $\kappa_0$ to Figure 6 could determine if $f$ harbors a hidden contribution of bending rigidity to gel elasticity[62].

Second, Figure 3 indicates that aspect ratio effects become progressively more pronounced as volume fraction decreases. The slower growth in $G'$ with $\phi$ for $r < 1$ (as observable from the aspect ratio dependence of $\Pi$) implies that below a specific transition concentration – which for this system we would extrapolate to be about $\phi \sim 0.2$ – $G'$ of discoid gels will always be higher than that of spheres. This observation coincides with prior results for rod gels[12]. The transition concentration likely coincides with the transition to fractality in the gels, which earlier has been discussed to occur at around $\phi \sim 0.1$ and below[63]. Identifying this transition concentration more specifically for anisotropic colloidal gels would identify the design space in which the multiplicate factors identified in this paper are available to formulators to achieve particular values of gel elasticity.

Third, although this work has addressed linear rheology, other non-linear rheological quantities – especially the yield stress and yield strain – warrant study. In their modeling, Shih et al show that the multiplicative factors reported in this study also impact yield rheology[19,24]. For instance, the gel yield stress is given to be proportional to $\kappa_0 \phi^{(2+\epsilon)/(3-d_f)}$; from Figures 4 – 6, at a characteristic concentration of $\phi = 0.01$, this dependency suggests that the yield stress in $r =$ 0.09 discoid gels would be at least 20-fold higher than that of sphere gels. This observation is consistent with measurements for boehmite rod gels, where, increasing the rod aspect ratio was found to generate as much as a factor of ~15 increase in the gel yield stress[12]. Additional study of non-linear rheological properties is warranted because they affect performance properties such paint application and formulation stability[2].

Fourth, by demonstrating how and why shape anisotropic particles can be self-assembled to produce highly elastic gels at low volume fractions, this study suggests new directions to design the rheology of gels. Although the present worked identified the multiplicative factors for the case discoids, these factors are generally applicable. For example, they could potentially be used to produce gels from particles of intricate shape that were selected to leverage the factors of hierarchical structure, anisotropic pair potential, and cluster compactness. Such structures could conceivably be more efficient at generating elasticity at low volume fractions than the discoidal shape studied here. This study's identification of the three independent, multiplicative factors therefore opens directions for gel design that are more specifically grounded in the physics of fractal structure, anisotropy, and self-assembly than previously available.

**Methods**

**Colloidal Particles**

The particles used in this study are polystyrene colloids. Fluorescently labeled carboxylate modified microspheres with diameter 0.98 ± 0.02 μm (F8821 FluoSpheres, Thermo Fisher Scientific, zeta potential = -52.4 ± 3.2 mV) were used for confocal microscopy experiments. Rheology experiments were performed using non-fluorescent, sulfate modified latex particles with diameter 1.00 ± 0.02 μm (S37498, Invitrogen, zeta potential = -55.6 ± 1.8 mV). The particle diameters were computed from scanning electron microscopy (SEM) images (TESCAN RISE, Michigan Center for Materials Characterization).

**Thermomechanical squeezing of PS spheres**

Colloidal discoids are generated by uniaxial compression of precursor polystyrene spheres. Both the fluorescent and non-fluorescent precursor spheres are subjected to the same treatment. The method is adapted from Ahn et al[16] and Hsiao et al[21]. The stock PS particles are thrice washed in de-ionized water and gently mixed with 10 wt% polyvinyl alcohol (PVA, molecular weight = 30 – 70 kDa, Sigma Aldrich) solution prepared in DI water. The choice of PVA and its concentration is based on prior work[64]. The mixture is then poured onto 35 mm petri dishes (Thermo Scientific) and allowed to dry for 24 hours at 25°C on a precision leveling platform (TrippNT). After the edges are trimmed, the dried PVA films with embedded spheres are sandwiched between silicone rubber sheets (50A durometer, 0.5mm thick, McMaster Carr) and placed in between two 15.2 cm x 15.2 cm stainless steel panels (1 mm thick, McMaster Carr). See Supplementary Figure 1 for the stack arrangement. The composite is then placed between heated platens of a bench-top press (Carver, Inc.) at 120°C (above the glass transition temperature of polystyrene, 90°C). The residence time is set to be 20 minutes for the films to reach the setpoint temperature (determined using a thermocouple thermometer, Fisher Scientific). A uniaxial compression is then applied to deform the film in which the spheres are embedded. The force is held for 20 min, following which the heat is turned off. The pressed films are allowed to cool under pressure to room temperature. The forces applied in this study are 750, 2800, 4200 kPa.

**Retrieving colloidal discoids**

Discoids are retrieved from the pressed PVA films following the procedure of Madivala et al[11]. The same procedure is followed for discoids prepared for both microscopy and rheology experiments. The PS-PVA films are dissolved in a 7:3 mixture of deionized water – isopropanol at 35°C for twelve hours with vigorous stirring. The solution is then heated to 60°C for 30 min to dissolve the PVA completely. The solution is centrifuged, and the recovered particles are thrice washed in the same solvent. The particles are then dispersed in deionized water and heated to 60°C for 30 min with vigorous stirring to dissolve any final traces of PVA. Finally, the particles are thrice washed in DI water. The concentration of particles is measured using a hemocytometer (NanoEnTek Inc.). The average zeta potential of the discoids is -45.3 ± 3 mV, indicating a stable suspension, similar to the seed spheres, whose zeta potential is -52.4 ± 3.2 mV (Zetasizer Nano ZSP, Malvern Instruments). Aspect ratio, $r$, and discoid major axis $2A$ are measured from images acquired by SEM of dilute samples in which discoids lie flat on the substrate. The minor axis *2B* is then obtained from conservation of volume of the seed sphere as per Hsiao et al[21].

**Gelation of colloidal particles for microscopy and rheology**

For gelation studies, the particles are redispersed in a buoyancy-matching mixture of deuterium oxide (151882, Sigma-Aldrich) and deionized water (resistance 18.2 MΩ, using Thermo Scientific DI Purifier) to prevent sedimentation effects [7]. Colloidal gels are assembled by the addition of $MgCl_2$ (68475, Sigma-Aldrich), which initiates aggregation[20]. We follow the 50-50 mixing rule where equal parts of particles in $H_2O$-$D_2O$ are mixed with equal parts of $MgCl_2$ solution to yield a final solution of desired particle volume fraction and [$MgCl_2$] = 10mM.

**Confocal microscopy of colloidal gels**

An inverted confocal laser scanning microscope (CLSM) (Nikon A1Rsi, equipped with NA = 1.4, 100x objective, oil-immersion type) is used to image the 3D microstructures of the gels self-assembled from fluorescently labeled particles. After the addition of $MgCl_2$ solution, the suspension is briefly and gently mixed for homogeneity and loaded into a 16-well chambered cover glass (Grace Bio-Labs, CultureWell, ChamberSLIP 16) mounted on the microscope stage above the objective. The chamber was closed to prevent evaporation. The gels form quiescently for 45 minutes before imaging. For visualization and microstructure characterization, 3D image volumes of size 512 x 512 pixels with pixel size 0.083 μm were acquired, beginning at the coverslip. The image stacks comprised of ~ 200 slices spaced at 0.083μm (acquired using Nikon AI Piezo z-drive). While acquiring image volumes, the intensity gain was gradually increased in the z-direction in steps of 1 unit for every 0.5 μm to compensate for the loss of image intensity at depths greater than 8 µm due to the refractive index mismatch between polystyrene particles and $H_2O$-$D_2O$ solution. CLSM visualization of the purified discoids are observed to be free of self-aggregation prior to the start of gelation (Supplementary Figure 4).

**Box-counting image analysis to compute fractal dimension**

The fractal dimension, $d_f$, of each gel network is computed from confocal micrographs using the box counting method implemented as a custom MATLAB program[45]. Raw CLSM images are subjected to a thresholding filter of value S to distinguish foreground and background pixels. Segmented image volumes are then divided into cubes of dimension $L$ x $L$ x $L$ (pixel³) (c.f. Figure 4(a))[47]. The value of $L$ is systemically varied from $L = 2$ pixels to $L = 64$ pixels. At each step, the number of cubes ($N(L)$) needed to cover all the foreground pixels corresponding to the gel is counted. Gel $d_f$ is then obtained from the power-law fit $N(L) = c_1(L/I)^{-d_f}$. Here, $I$ is the image size (in pixels).

To address the S dependence inherent to this method, we apply the criteria of Thill and co-workers[45]. First, we vary S from 0 to 255 and plot $d_f$ versus S as shown in Supplementary Figure 5. The optimal threshold – where $d_f$ does not change significantly – is identified by fitting the data to a third order polynomial and identifying the point where the concavity of the curve is zero. The optimal thresholds, found to be S = 136, 144, 152 and 141 for $r$ = 1, 0.42, 0.20 and 0.09 respectively, were then used to compute the $d_f$ values reported in Figure 4. This analysis accounts for any differences in instrument settings for image acquisition used due to batch-to-batch variation in particle fluorescence intensities.

To test the fidelity of this method, a complementary approach to identify the optimal threshold was followed. First, the $d_f$ of sphere gels was determined from their radial distribution function, $g(r_0)$ [35]. Here, $g(r_0)$ is computed from particle centroids using tools available in TRACKPY[65] and FREUD[66] Python libraries. $g(r_0)$ describes the average number of particles at a distance $r_0$ from a basis particle, relative to that of an ideal gas. For fractal structures, $g(r_0) \sim r_0^{d_f-3}$ for $r_0/2A > 3$ [36]. Following this equation, the fractal dimension is obtained from the slope of the log-log plot of $g(r_0)$ data (Supplementary Figure 6). This method yields $d_f = 1.89 \pm 0.04$ for sphere gels. Second, this value is matched with Supplementary Figure 5 to find the threshold at which the $d_f$ values for sphere gels match. A value of S = 138 is identified, which is consistent with the numbers identified above. The $d_f$ for all the discoid gels were then computed at this threshold value. Across the range of aspect ratios studied, the relative standard deviation in $d_f$ computed using the two methods is less than 2.5%, thus validating the box-counting method.

**Skeletonization image analysis to compute backbone topography**

The backbone of sphere and discoid gels was extracted from confocal image volumes following the method of Immink et al[51] and Kollmannsberger et al[53]. Raw confocal images were first pre-processed by applying a standard Gaussian blur (cubic kernel of size 5 voxels) to filter out noise and subsequently binarized using Otsu's method[67]. Morphological closing was then performed (cubic structural element of size 1 voxel) that removed discretization noise followed by the removal of small unconnected structures (components of volume < 1% of the largest connected component in the image). These processes were implemented using the 3D Volumetric Image Processing Toolbox available in MATLAB. The pre-processed image volumes were then skeletonized using the medial axis thinning algorithm, available through the *bwskel* function in MATLAB to produce a voxel thick backbone of the gel network[68]. The backbone dimension, $d_B$ and anisotropy, $\epsilon$ were then computed as per the following post-processing steps. The skeleton was converted into a non-directed weighted graph of nodes and edges using the open-source image-processing add-on to MATLAB, *Skel2Graph3D*[69]. The length, $l$ and coordinates of the shortest connected path between all node pairs were then extracted using Dijkstra's algorithm available in MATLAB. Euclidian distances, $r_E$ between node pairs were directly computed from their coordinates on the skeleton. Coordinates of the shortest path was then used to compute their center of mass radius of gyration, $r_g$ as well as their radius of gyration projected onto a plane perpendicular to their end-to-end axis, $r_\perp$. Then, $d_B$ and $\epsilon$ were obtained through fits to the measured data as $l = c_2 r_E^{d_B}$ and $r_\perp^2 = c_3 r_g^{2\epsilon}$. [42,54–56]

**Rheological characterization of colloidal gels**

Rheological measurements are performed with a stress-controlled DHR-3 rheometer (TA Instruments) using a 40 mm stainless steel parallel plate geometry and a Peltier temperature-controlled plate (TA Instruments). Roughness 600 grit sandpaper (Part # 47185A51, McMaster Carr) was attached to both the top and bottom geometry surface to prevent wall slip[70]. The sample gap was set to be 500 μm. The temperature for all the rheology measurements is set to 20°C. Suspensions were loaded onto the Peltier plate, the top geometry was lowered, and particles were allowed to quiescently assemble at the measurement gap for a gelation time of 45 minutes. An insulated solvent trap cover (TA Instruments) was used to prevent evaporation. Gel rheology was then measured by performing oscillatory strain amplitude sweeps ranging from $\gamma$ = $10^{-4}$ to $10^{-1}$ at a constant frequency of 1 rad/s.

We checked that the above choice of sandpaper and gap addressed any potential wall slip and confinement effects by performing measurements for sphere gels using different roughness sandpapers and geometry gaps (Supplementary Figure 7). The relative standard deviation of these measurements for linear storage modulus, $G'$ is less than 10% indicating the results are independent of choice of measurement setup[22]. Instrument sensitivity limits on the strains sweep plots (Figure 2) were determined by performing experiments with PEO (molecular weight ~ 1 x $10^6$ g/mol, Sigma Aldrich) solutions at 2.0, 2.5, 3.0 and 4.0 wt% to determine the lower stress limits of the rheometer (Supplementary Figure 8).

Rheology: How to Avoid Bad Data. in *Complex Fluids in Biological Systems* (ed. Spagnolie, S. E.) 207–241 (Springer, 2015).

**Acknowledgements**

Support for this study was provided by the National Science Foundation (No. CBET-1702418). For scanning electron microscopy, the authors acknowledge the financial support of the University of Michigan College of Engineering and NSF grant #DMR-1625671, and the technical support from the Michigan Center for Materials Characterization.

**Author Contributions**

P.K.K. and M.G. synthesized the colloidal particles and carried out the experiments. P.K.K., M.J.S. and M.G. designed the project, analyzed the data and wrote the manuscript. All authors discussed and commented on the manuscript.

**Competing Interests Statement**

The authors declare no competing interests

**Materials & Correspondence**

Correspondence and requests for materials should be addressed to M.J.S. (email: [mjsolo@umich.edu](mailto:mjsolo@umich.edu)) and M.G. (email: [maheshg@umich.edu](mailto:maheshg@umich.edu))

**Figures**

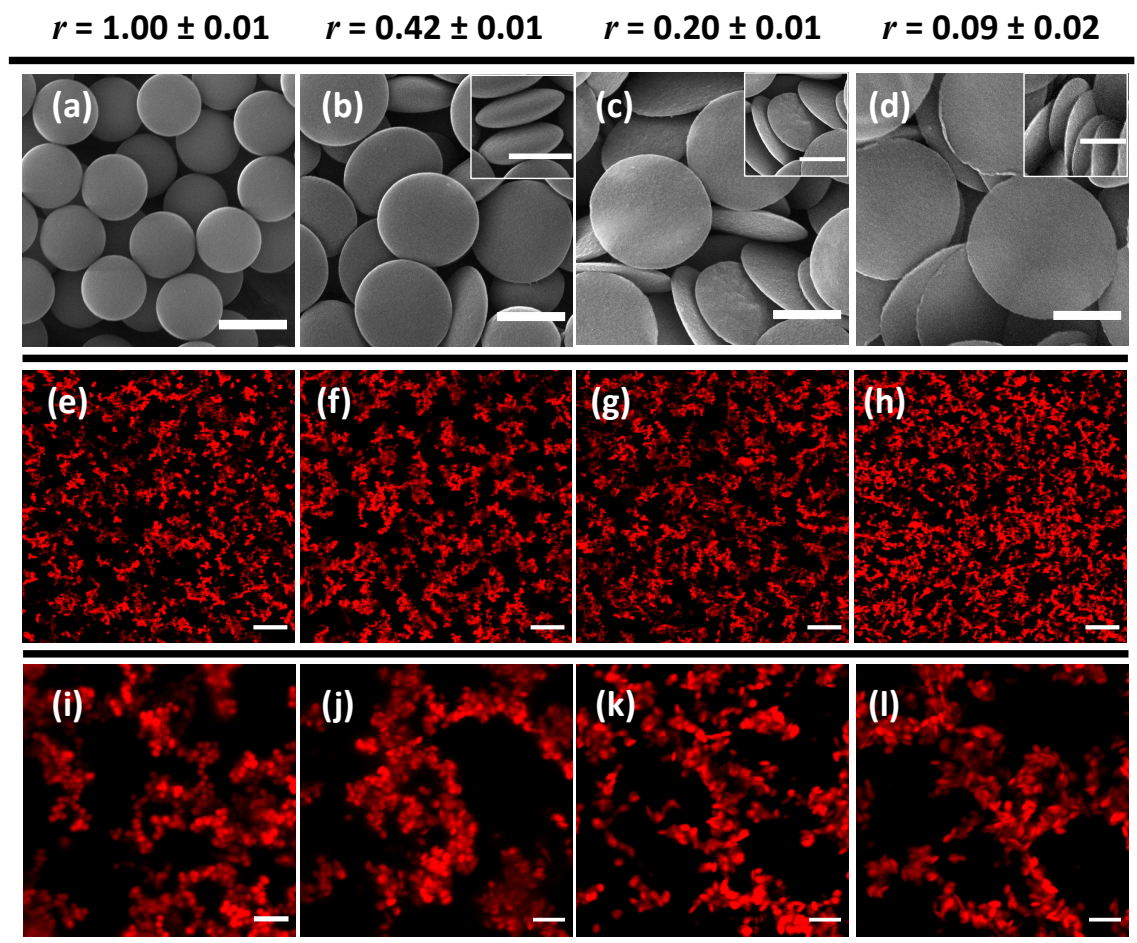

**Figure 1. Self-assembly of sphere and discoidal gels.** (a-d) Representative SEM images of colloidal particles with aspect ratio $r$ = 1, 0.42, 0.20, and 0.09 respectively. Insets of different viewing angles show the thickness of the discoids. Scale bars for (a-d) are 1μm. Confocal projections ($\Delta z = 20 \mu m$) of colloidal gels with aspect ratio $r$ = (e, i) 1.0, (f, j) 0.42, (g, k) 0.20, and (h, l) 0.09. Here, $\phi = 0.015$ and [$MgCl_2$] = 10 mM. Scale bars are (e-h) 20 μm and (i-l) 5 μm.

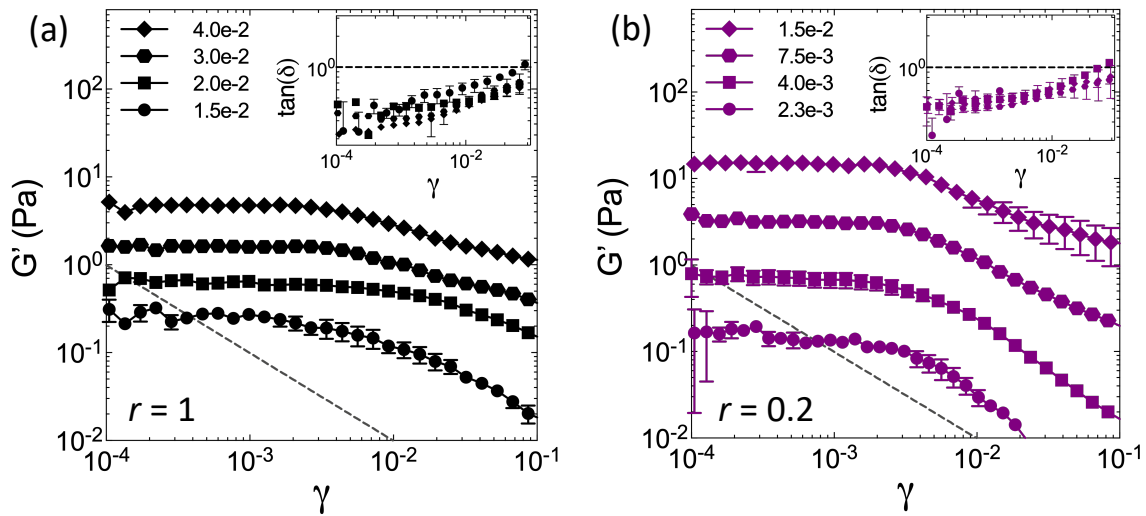

**Figure 2**. **Shear rheology of sphere and discoid gels.** Storage modulus ($G'$) and (inset) tan($\delta$) as a function of strain amplitude, $\gamma$ at $\omega = 1$ rad/s of colloidal gels made from (a) spheres ($r = 1$) and (b) discoids ($r = 0.20$). Dotted line indicates the instrument sensitivity limits (c.f. Methods). Corresponding strain sweep measurements for $r = 0.42$ and $r = 0.09$ discoids are included in Supplementary Figure 2.

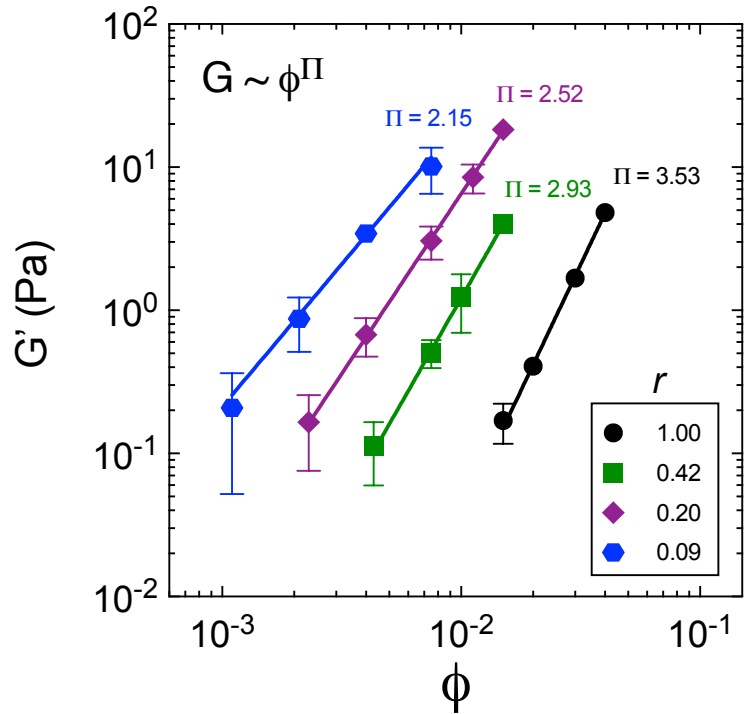

**Figure 3**. **Gel elastic modulus as a function of particle volume fraction and aspect ratio.** The dependence of linear storage modulus, $G'$ on particle volume fraction $\phi$ for colloidal gels made from spheres ($r = 1$) and discoids ($r = 0.42$, $0.20$ and $0.09$). The reported $G'$ is an average over the linear region which is identified as the strain amplitudes less than the point at which $G'$ deviates by 5% from its maximum value[19]. The uncertainties in the exponents are $\pm 0.15$, $\pm 0.05$, $\pm 0.10$ and $\pm 0.16$ for $r = 0.09$, $0.20$, $00.42$ and $1.00$ respectively. At some data points, error bars are not visible because they are small relative to the marker size.

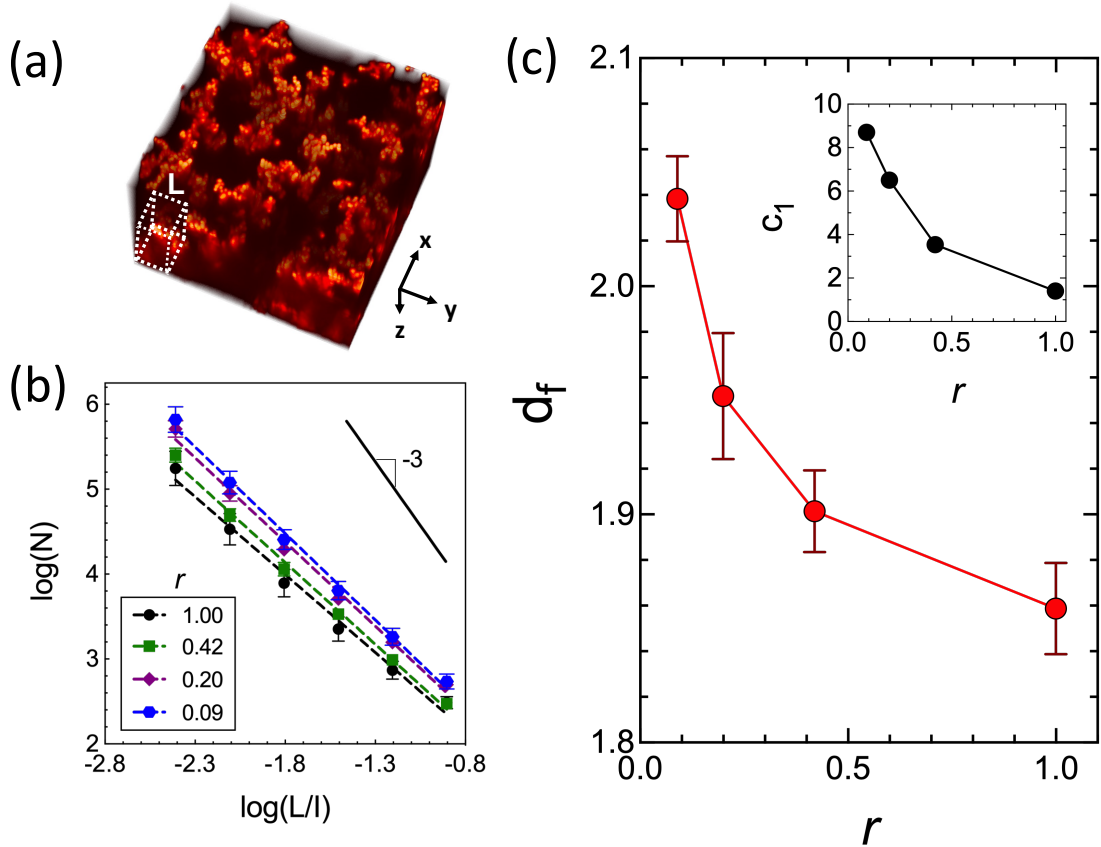

**Figure 4. Characterization of network fractal dimension as function of particle aspect ratio.** (a) 3D confocal microscopy visualization of a sphere colloidal gel to characterize fractal microstructure. Here, $\phi = 0.015$ and $[MgCl_2] = 10$ mM. The white dotted lines illustrate a single cube in the box counting method. $L$ is the resolution of the cube. (b) Log-log plot of number of cubes ($N$) vs. dimensionless cube size ($L/l$) for colloidal gels made from spheres ($r = 1$) as well as $r = 0.42$, 0.20 and 0.09 discoids. The slope of the curve is the fractal dimension of the aggregates. The solid line represents a Euclidean scaling of $N \sim L^{-3}$. (c) The gel fractal dimension, $d_f$ as a function of aspect ratio $r$. (inset) Prefactor $c_1$ as a function of $r$. The lines are included to guide the eye. Error bars for $c_1$ are small relative to the marker size.

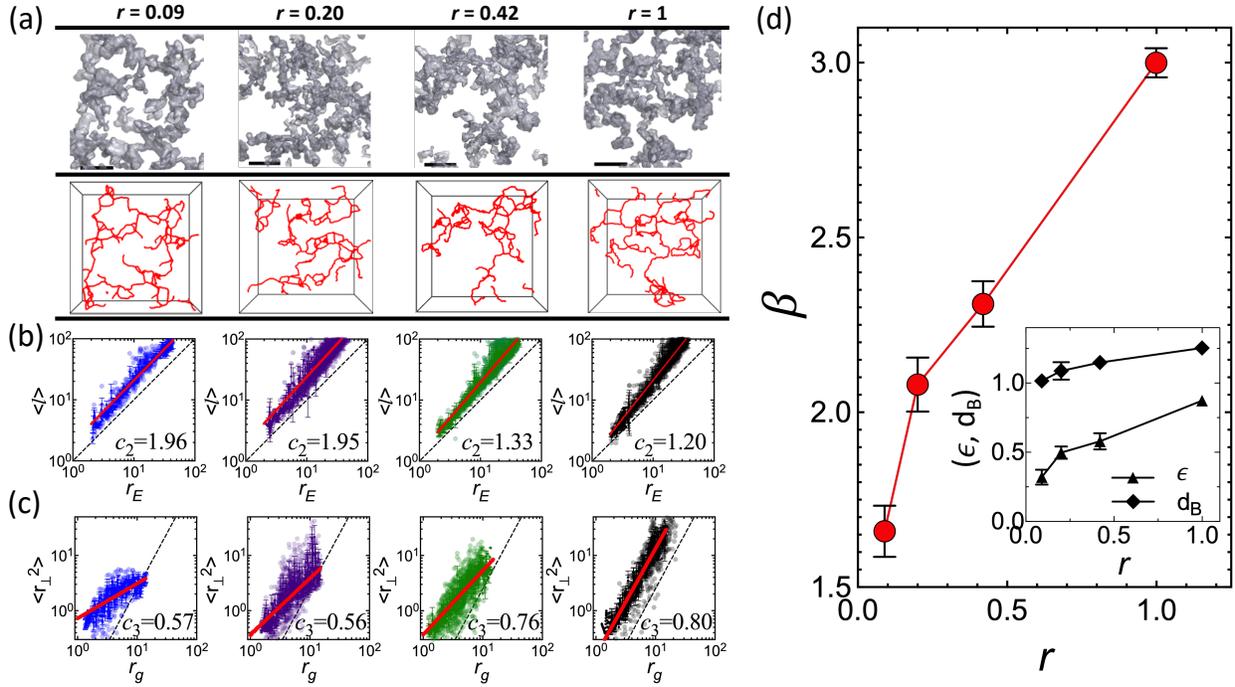

**Figure 5**: **Characterization of backbone topography as function of particle aspect ratio.** (a) Representative iso-surface rendering of binarized image volumes of the gels (top row) and their corresponding backbone extracted using skeletonization (bottom row). The aspect ratios of the constituent particles are indicated on the top. Corresponding real-space confocal micrographs are shown in Supplementary Figure 9. (b) Log-log plot of $\langle l \rangle$ versus $r_E$ for different aspect ratio gels to determine the backbone fractal dimension $d_B$. (c) Log-log plot of $\langle r_\perp^2 \rangle$ versus $r_g$ for different aspect ratio gels to determine backbone anisotropy, $\epsilon$. The solid lines in (b) and (c) are respectively power-law fits: $\langle l \rangle = c_2 r_E^{d_B}$ and $\langle r_\perp^2 \rangle = c_3 r_g^{2\epsilon}$. The dashed lines indicate the case (b) $d_B$=1.00 and (c) $\epsilon$ =1.00. Values of the pre-factors $c_2$ and $c_3$ are included in the inset. (d) Aspect ratio dependence of the elasticity exponent, $\beta$ and (inset) aspect ratio dependence of $d_B$ and $\epsilon$. Lines in (d) are drawn to guide the eye.

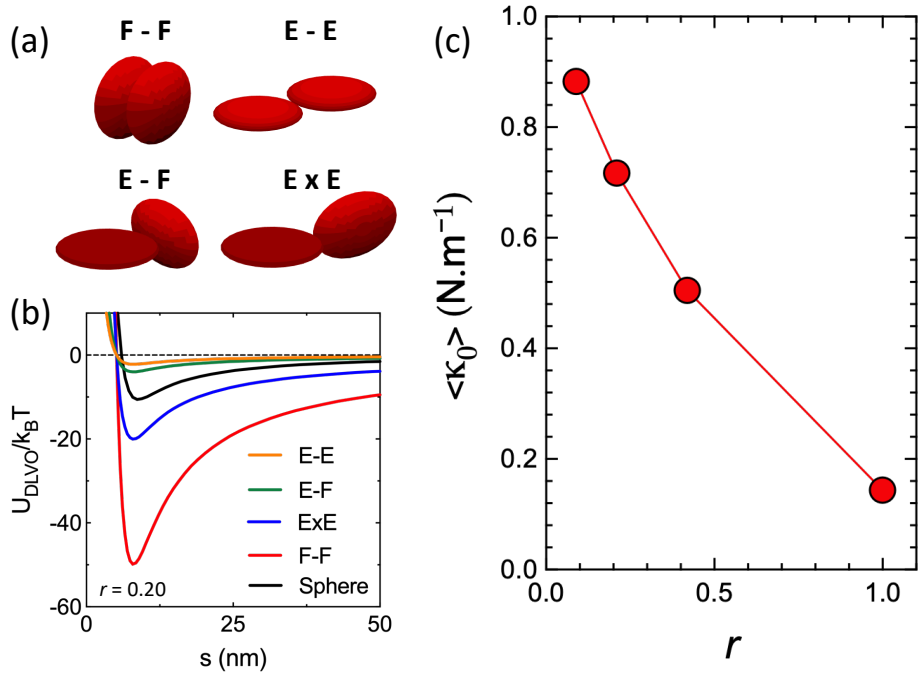

**Figure 6. Interparticle bond strength as function of particle aspect ratio.** (a) Limiting cases of the pairwise orientation distribution function of discoids: face-to-face (F-F), edge-to-edge (E-E), edge-to-face (E-F) and edge-on-edge (ExE). Discoids are not drawn to represent any specific aspect ratio used in this study. (b) DLVO pair-potential, $U_{DLVO}$ normalized by $k_B T$ for sphere pairs and discoid pairs corresponding to relative orientations shown in (a). Discoid pair-potentials are for the case $r = 0.20$. (c) Aspect ratio dependence of the orientationally averaged bond spring constant $\langle \kappa_0 \rangle$. Lines are drawn to guide the eye.

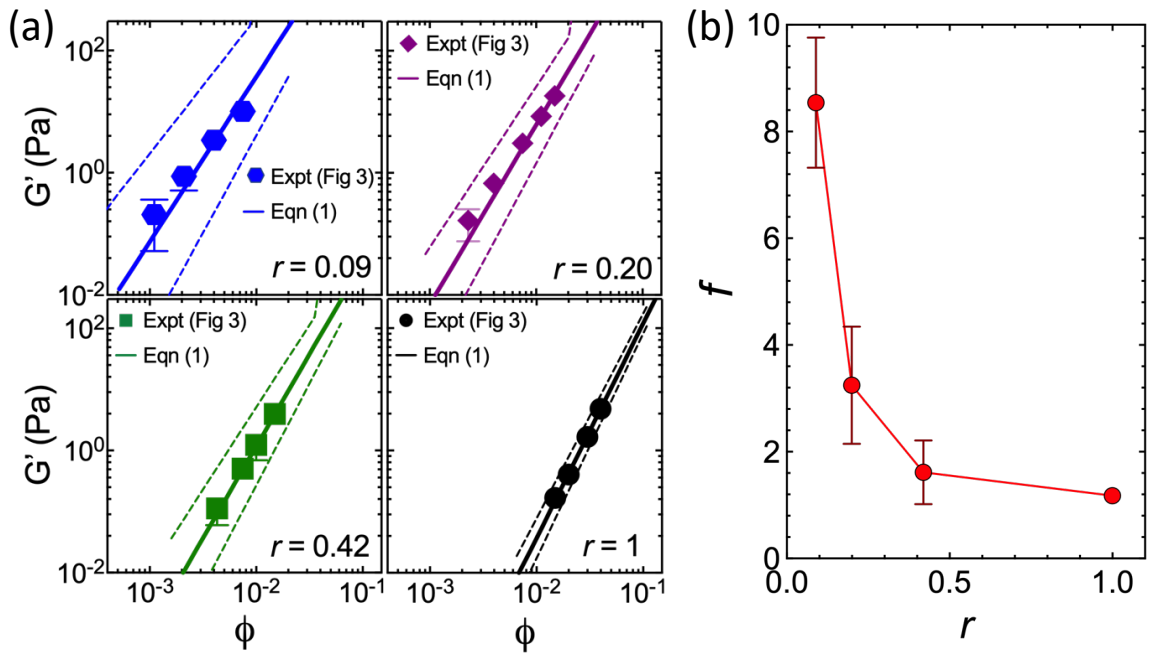

**Figure 7. Predicting aspect ratio dependence of gel elastic moduli.** (a) Prediction of gel elastic modulus as function of volume fraction for sphere ($r = 1$) and discoid gels ($r = 0.09, 0.20$ and $0.42$). The solid line represents predictions using equation (1) and dashed lines are confidence intervals obtained by propagation of errors in measured model parameters. (b) Aspect ratio dependence of proportionality constant, $f$.

# Supplemental Information – Shape anisotropy enhances the elasticity of colloidal gels through mechanisms that act multiplicatively

Peng-Kai Kao[†], Michael J. Solomon[†,*], and Mahesh Ganesan[†,*]

[†]Department of Chemical Engineering, University of Michigan, Ann Arbor, Michigan

*Corresponding Authors:

Prof. Michael J. Solomon

Address: North Campus Research Complex, Building 10 – A151, 2800 Plymouth Road, Ann Arbor, MI 48109

Phone: 734-764-3119

Email: mjsolo@umich.edu

Dr. Mahesh Ganesan

Address: North Campus Research Complex, Building 20 –106W, 2800 Plymouth Road, Ann Arbor, MI 48109

Phone: 734-546-0210

Email: maheshg@umich.edu

**Supplementary Figures**

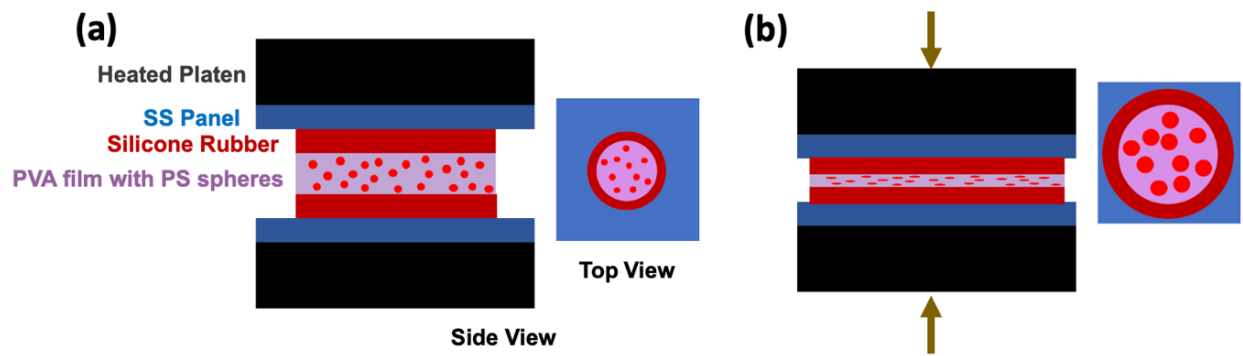

**Supplementary Figure 1. Schematic of discoid synthesis process.** (a) Seed polystyrene (PS) microspheres, embedded in a polyvinyl alcohol (PVA) film, are sandwiched between silicone rubber carriers (50A durometer, 0.5mm thick) and placed in between 1 mm thick stainless-steel (SS) panels (15.2 cm x 15.2 cm) housed inside a heated plate press. (b) Uniaxial compression leads to equi-biaxial deformation of the layers resulting in oblate discoids[1,2].

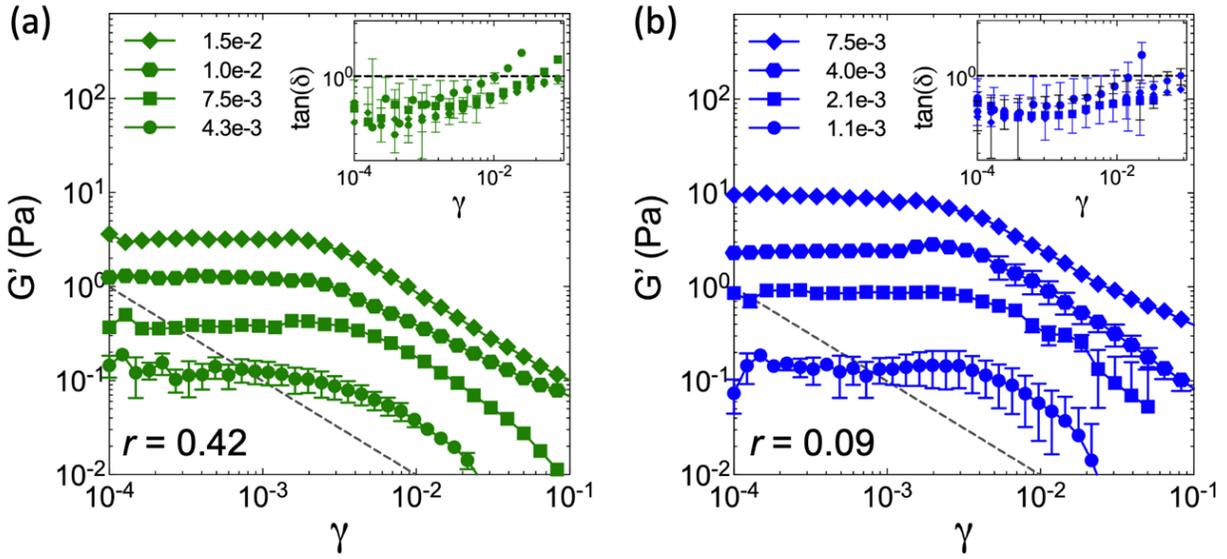

**Supplementary Figure 2. Shear rheology of discoid gels.** Storage modulus (G') and tan(δ) (inset) as a function of strain amplitude (γ) for colloidal gels made from discoids of aspect ratio (a) $r = 0.42$ and (b) $r = 0.09$. The oscillatory frequency is $\omega = 1$ rad/s. The dashed lines represent the instrument sensitivity limits. Concentration of $MgCl_2$ is 10mM. Different curves correspond to different volume fractions.

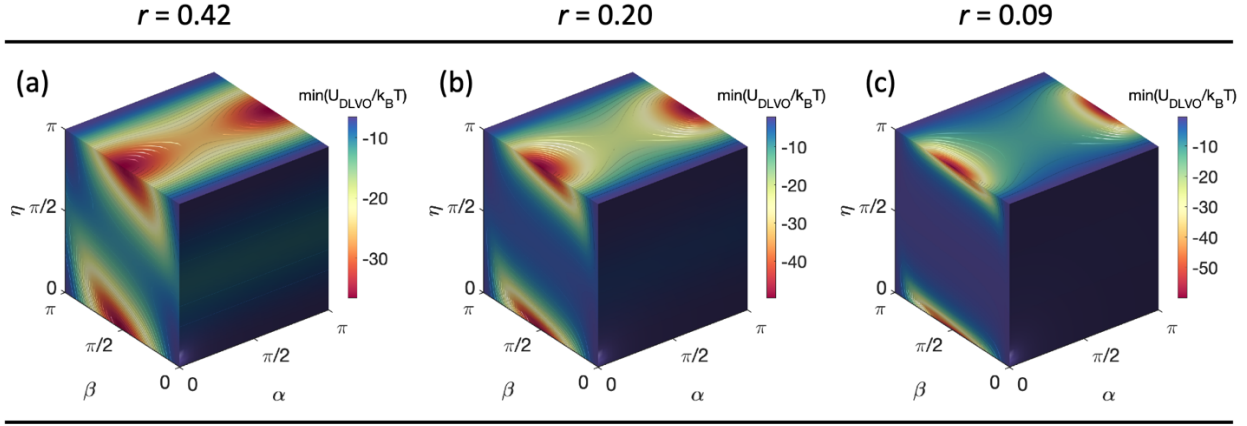

**Supplementary Figure 3. Potential energy for interacting discoid pairs.** Strength of the DLVO interparticle potential energy ($U_{DLVO}$) as a function of relative orientation for a pair of discoids with aspect ratio (a) $r = 0.42$, (b) $r = 0.20$, (c) $r = 0.09$. The DLVO potential here is a sum of nonretarded van der Waal's attraction and electrostatic repulsion potential. The expressions for calculating the orientation dependent DLVO potentials for discoids are given in Schiller et al[3]. They describe the mutual orientation of a pair of discoids in terms of angles $\alpha, \beta$, and $\eta$[3]. Here, $\alpha$ denotes the angle of relative orientation of the two coordinate systems that define the principal radii of curvature of the two particles, $\beta$ is the polar angle relative to the major axis and $\eta$ is the angle enclosed by their axes of revolution. This figure highlights the range of potential energies possible due to interaction anisotropy accorded by the discoidal particle shape. The energetically favorable F-F configuration discussed in the main text corresponds to $(\alpha, \beta, \eta) = (0, \pi/2, 0)$. We find that the F-F bond energy increases with increasing discoid anisotropy (decreasing $r$). The E-E, E-F and ExE configurations shown in Figure 6(a)-(b), respectively correspond to $(\alpha, \beta, \eta) = (0,0,0)$, $(0,0, \pi/2)$ and $(\pi/2, \pi/2, 0)$. To compute the bond spring constant, we average over all the possible potential states plotted in (a), (b) and (c) by considering the probability for sampling the different orientations to satisfy the Boltzmann

distribution[4]. Here, the potential energy is normalized by $k_BT$, where $k_B$ = Boltzmann constant and T is the temperature.

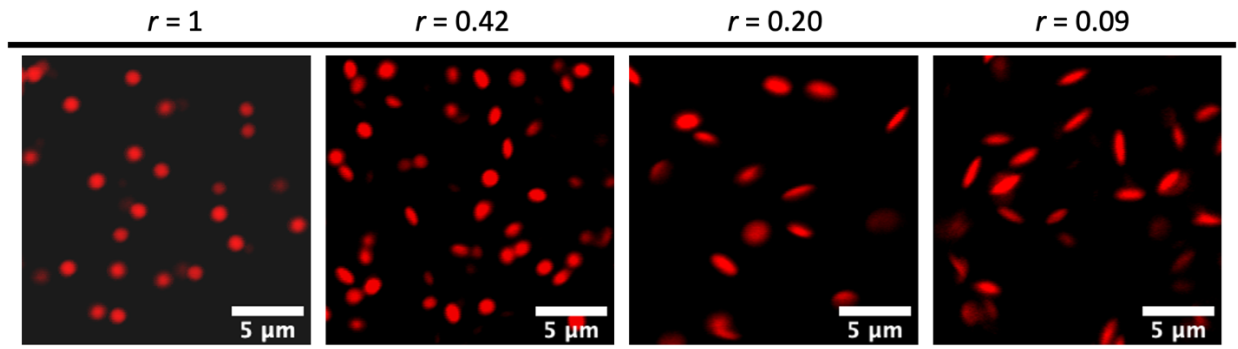

**Supplementary Figure 4. Dilute discoidal suspensions are well-dispersed prior to gelation.** Confocal micrograph of dilute suspension of spheres ($r = 1$) and synthesized discoids ($r = 0.42$, 0.20 and 0.09).

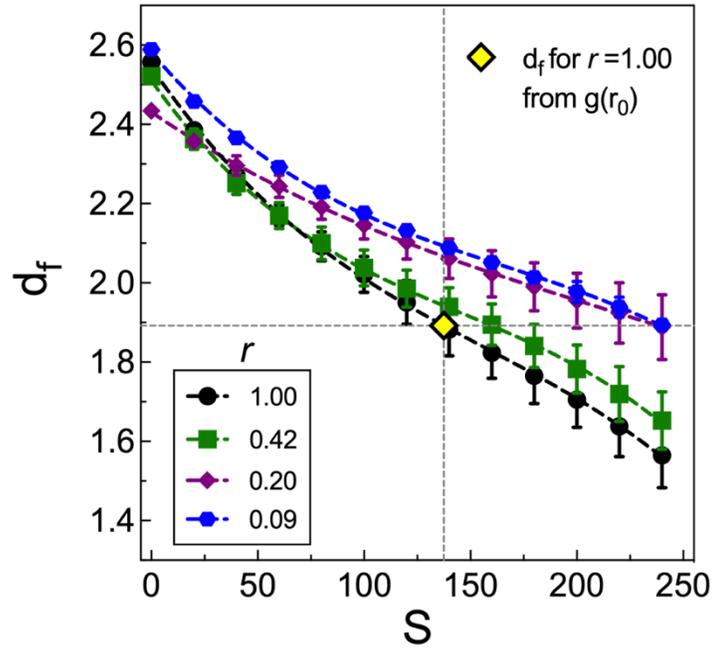

**Supplementary Figure 5.** Dependence of box-counting fractal dimension, $d_f$, on image threshold parameter, S, measured from confocal image volumes of colloidal gels consisting of spheres ($r = 1$) and discoids ($r = 0.42$, 0.20 and 0.09). The optimal threshold derived from fitting, as described in the main text, for each aspect ratio, are S = 136, 144, 152 and 141 for $r = $ 1, 0.42, 0.20 and 0.09 respectively. The $d_f$ for sphere gels ($r = 1$) obtained from its radial distribution function, $g(r_0)$, is indicated by the horizontal dashed line and marked as solid diamond on the curve; the corresponding threshold, S = 138, is indicated by the vertical dashed line.

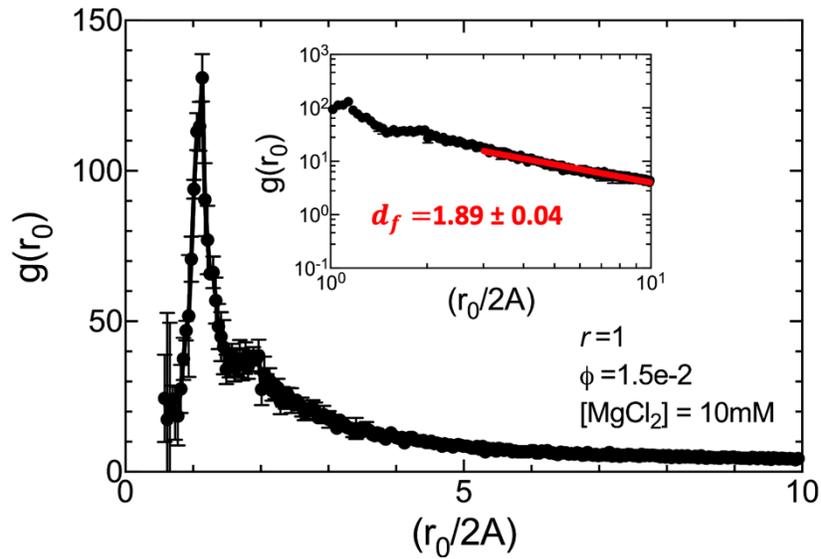

**Supplementary Figure 6.** Radial distribution function, $g(r_0)$ of sphere gels ($r=1$). In the fractal regime, the functional dependence of $g(r_0)$ is given as: $g(r_0) \sim r_0^{d_f-3}$ [5]. As shown in the inset plot, we observe this fractal scaling for $r_0/2A > 3$. The solid line is a fit to the data as per the above equation and the resulting $d_f$ is mentioned. The range selected for fractal analysis is consistent with the approach of Lattuada and co-workers[7]. The particle volume fraction is $\phi = 0.015$ and [MgCl$_2$] = 10mM.

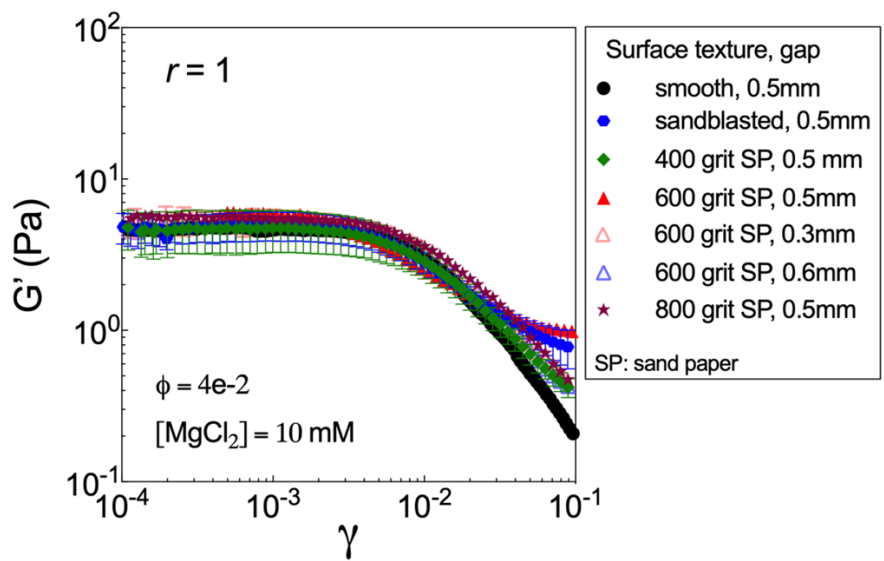

**Supplementary Figure 7. Checking for slip in oscillatory shear rheology**. Storage modulus (G') as a function of strain amplitude ($\gamma$) for a sphere gel ($\phi$ = 0.04, [MgCl$_2$] = 10mM) measured using a 40mm diameter parallel plate geometry having different surface textures and at different measurement gaps. When testing with sandpaper, the sandpaper was attached to both the bottom and top bounding surfaces. The relative standard deviation in the linear storage modulus is less than 10% indicating the upper bond on any potential effects of wall-slip [8,9].

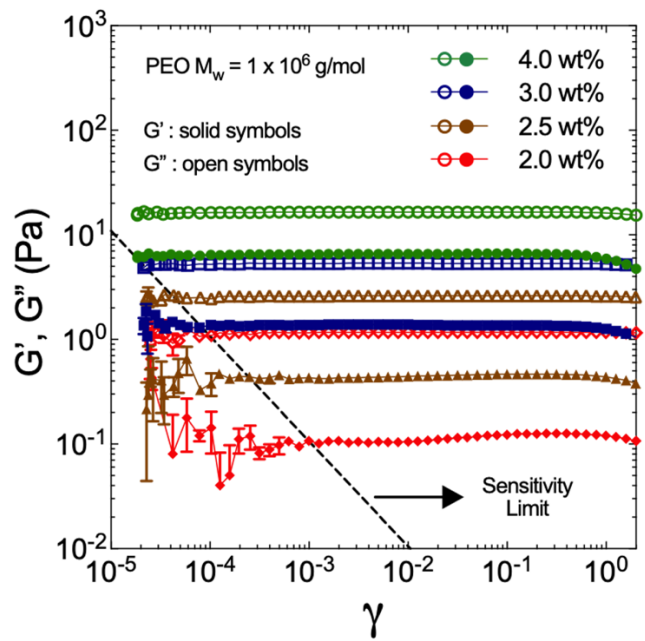

**Supplementary Figure 8. Rheometer sensitivity limit.** The lower stress limit of the rheometer used in this study was estimated from oscillatory amplitude sweep measurements of polyethylene oxide (PEO) solutions of different concentrations (2.0, 2.5, 3.0 and 4.0 wt %). Shear amplitudes at which the measured storage (G') and loss (G") moduli were consistent – having a variation of less than 2.5% with increasing shear strain – define the lower stress limit of the rheometer, marked here as the dashed line.

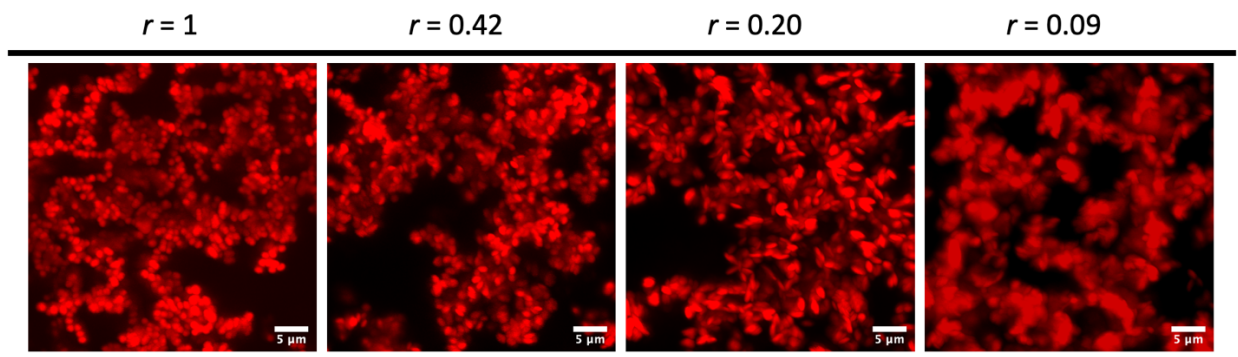

**Supplementary Figure 9. Confocal projections for skeletonization analysis.** Real space confocal maximum projections ($\Delta z = 20\mu m$) of colloidal gels consisting of spheres ($r = 1$) and discoids ($r = 0.42, 0.20$ and $0.09$) whose iso-surface rendering and extracted skeletons are shown in Figure 5(a) of the main manuscript. Here, $\phi = 0.015$ and $[MgCl_2] = 10mM$.